# The Effects of GQM+Strategies on Organizational Alignment

*Jürgen Münch[1], Fabian Fagerholm[1], Petri Kettunen[1], Max Pagels[1], Jari Partanen[2]*

[1]University of Helsinki, Department of Computer Science
P.O. Box 68, FI-00014, Helsinki, Finland
[2]Elektrobit Corporation, Wireless Business Segment
Tutkijantie 8, FIN-90690, Oulu, Finland

[1]firstname.lastname@cs.helsinki.fi, [2]jari.partanen@elektrobit.com

*Abstract*

*The increasing role of software for developing products and services requires that organizations align their software-related activities with high-level business goals. In practice, this alignment is very difficult and only little systematic support is available. GQM+Strategies is a method that aims at aligning organizational goals, strategies, and measurements at all levels of an organization in a seamless way. This article describes a case study of applying GQM+Strategies in a globally operating industrial R&D organization developing special-purpose device products for B2B customers. The study analyzes how GQM+Strategies has helped clarify and harmonize the goal set of the organization. Results of the study indicate improved alignment and integration of different goals. In addition, the method helped to make the initially informal goal-setting more transparent and consequently enabled revising it while new, more important goals were discovered and comprehended. Moreover, several elements affecting the achievement of goals as well as impediments were identified.*

## 1 Introduction

Software plays an enormous role for companies in developing competitive products and services. The digital transformation of most industry sectors requires the effective and efficient integration of software capabilities in all kinds of products and services as a prerequisite for building successful business models and in order to safeguard a future place in the market. In consequence, managers in organizations need to understand how to use software and how to link software-related activities with higher-level goals such as business goals. Without such kind of understanding, it will be increasingly difficult to acquire or

Jürgen Münch, Fabian Fagerholm, Petri Kettunen, Max Pagels, Jari Partanen

retain a market share. In addition, resources are limited and it is key for most organizations to avoid activities that do not contribute to overall goals and the success of an organization. In order to avoid wasting time, effort, and money, it is highly necessary to come up with the right goals and business strategies and to guarantee that the organization is aligned towards these goals and strategies. This refers essentially to quickly identifying successful products or services and to having effective mechanisms in place to link all activities of an organization towards these goals and strategies. While avoiding the creation of wrong products and services is addressed by approaches such as Lean Startup or Customer Development, systematic support for linking goals and strategies throughout the whole organization is widely missing. The need for such support is recognized by a multitude of studies in the literature. Pfeffer, for instance, states that high-performing organizations are goal-oriented by nature [12]. A common, compelling direction aligns the organizational activities at different levels and functions even in distributed settings. Conversely, without known shared goals, different parts of larger enterprises in particular may not be able to contribute efficiently to the overall business success.

Visual controls are key management tools in Lean enterprises [8]. When the goals and current performance information is transparently available across the enterprise all the time, the different actors can agree on and steer their contributions coherently in collaboration even in globally distributed organizations. Measurements form an inherent part of organizational information. In order to be able to monitor the progress of goal achievement, each goal should be coupled with appropriate measures [14]. Ideally, fit and prompt measures can help achieving and maintaining alignment, and boost performance. It follows that there is a compelling need to be able set and refine transparently shared, mutually agreed goals (e.g., customer satisfaction) in all parts of distributed R&D organizations, and track their progress continuously. This allows for planning future directions in systematic ways, which in turn promises that strategic changes (even enterprise transformations) are more tractable and that the effects of implementing changes are more visible.

GQM$^+$Strategies is an approach for aligning, harmonizing, and communicating goals and strategies of an organization in order to direct the organization in the same direction [3]. It seamlessly integrates goal-oriented measurement in the alignment process and therefore allows to manage, control, analyze, and change goals and strategies based on data. This paper describes a case study of applying GQM$^+$Strategies in a globally operating industrial R&D organization. The study examined how the linkage between organizational goals changed over time in the organization by applying GQM[+]Strategies and in consequence how strategic alignment was achieved.

The rest of this paper is organized as follows. Section 2 gives an overview of GQM[+]Strategies. Section 3 describes related work. Section 4 describes the research approach of this study, including the context, the research method, the hypotheses, and the execution of the study. Section 5 describes the results of the study. Section 6 discusses the implications of the results and the limitations of the study. Finally, in Section 7, the findings are summarized and possible future work is outlined.

## 2  GQM[+]Strategies

GQM[+]Strategies [1] supports alignment throughout an organization by introducing relationships between goals and strategies through measurement. When a goal is defined, a strategy (i.e., a set of actions) is decided upon to achieve it. Strategies are then connected to more specific subgoals, each with their own set of sub-strategies. This forms an interconnected grid of goals, typically general in nature on the business level and more specific on lower organizational levels. It is this specific linkage between goals via strategies that supports alignment. A goal with no connected parents or children, for instance, is a sign that the goal does not contribute to any other part of the organization, and is thus unaligned. Applying GQM[+]Strategies promises the following benefits for organizations [9]:

1. Clarify goals
2. Harmonize goals
3. Align goals and strategies
4. Communicate business goals throughout the entire organization
5. Monitor the deployment of strategies
6. Obtain feedback about strategies and goals

Benefits 1) - 3) in particular pertain to goal alignment and are part of the case study presented in this paper; the end result of these activities is a set of strategic goals aligned in such a fashion that the satisfactory achievement of one goal has a positive impact on others. For the field of software engineering, this means sharing a common understanding of how software development can contribute to business goals and vice versa.

Applying GQM[+]Strategies in practice is an iterative process: changes in the grid of goals and strategies is to be expected before a well-aligned result is achieved [2].

## 3  Related Work

There are several approaches for supporting organizational alignment. Typically, they focus on specific aspects only or are highly domain-specific. COBIT, Practical Software and Systems Measurement, the Balanced Scorecard, and Six Sigma are examples of goal alignment approaches.

COBIT [6] is an approach for IT governance that not only distinguishes between goals on different levels, but also provides a predefined set of metrics against which the fulfillment of IT-specific goals can be measured. Predefined metrics offer a level of uniformity regardless of where COBIT is applied. Conversely, GQM[+]Strategies supports customizability when defining metrics.

Practical Software and Systems Measurement (PSM) [13] is an approach that can be used to collect, analyze and manage aspects of software and software systems. The characteristics and needs of a software project guide the selection of suitable metrics, providing a solution for measurement of predefined software-specific goals. In contrast to GQM[+]Strategies, PSM focuses on the project level and does not explicitly support the measurement of goals on higher organizational levels.

Balanced Scorecard (BSC) [5] is a strategic management tool that can be used to align the business activities and strategies of an organization and to monitor how an organization is performing against the goals it outlines. BSC addresses four perspectives (financial, customer, internal business processes, learning and growth) in order to derive measures on the higher levels of an organization. In contrast to GQM[+]Strategies, the linkage to lower levels of an organization including measurements and analytical reasoning is not systematically incorporated [7].

Six Sigma [11] is a quality improvement strategy. It is mainly used to identify and eliminate the causes of defects in business and manufacturing processes. Unlike GQM[+]Strategies, it is based on a quality paradigm that stems from production processes, i.e., repeatability of processes is assumed and minimizing variability of processes is seen as key to improving quality. This is not easily transferable to to software processes that are development processes by nature.

Assessing the effectiveness of a particular paradigm or approach in the field of software engineering is challenging: businesses and organizations rarely share the same goals, corporate structure or ways of working. As a result, the practical application of a goal alignment approach is likely to vary from organization to organization.

## 4 Case Study Approach and Execution

Our study can be characterized as a single-case study [15]. Case studies focus on understanding the dynamics of single settings [4]. In this case, we sought to understand whether and how the process of applying the GQM⁺Strategies approach would lead to a more aligned set of strategic goals for the case company. In this section, we describe the case company and the research problem, the decisions made with regard to application of case study method, and the execution of the study. In a previous paper [10], we have reported on experiences and lessons learnt during the application of the GQM⁺Strategies approach, and will therefore focus here on the parts that are most relevant for understanding the case study itself and its results.

### 4.1 Context and Problem

The case company Elektrobit operates in the automotive and wireless product and service industries. Elektrobit develops platform-based, integrated, special-purpose devices for its customers. The company's Wireless Business Segment, the branch under study, is responsible for devices with wireless capabilities, such as specialized Android-based handsets.

Initially, the improvement of organizational visibility was chosen as the primary focus of the application of GQM⁺Strategies. Shortly after, this was coupled with improving overall customer satisfaction. The scope of the goal alignment effort in this study encompasses these two themes.

Elektrobit was selected for this case study as the context of the company and its focus areas were ideal for assessing the practical effects of a goal alignment approach in the field of software engineering.

Table 1: Research Problem Setting.

| Element | This Investigation |
|---|---|
| Object | Development process |
| Purpose | Characterize and understand |
| Focus | Goal alignment |
| Perspective of | Platform Project Manager |
| Context | Distributed platform development |

## 4.2 Research Approach

The application of GQM$^+$Strategies consisted of a set of co-located planning meetings, followed by a series of workshops held via teleconference. A total of five workshops were held in order to elicit information and to develop a harmonized and aligned goal set for the target company. These workshops were attended by four researchers from the University of Helsinki and a variable number of representatives from Elektrobit. During each session, the specific goals of the company were discussed in detail and recorded in the form of notes and audio clips. After each workshop, a *grid* of goals and possible strategies was developed by the researchers. This grid was then validated and changed by company representatives in the following workshop.

Quantitative data derived from each grid revision during the course of the workshops with Elektrobit representatives serves as the primary data source for assessing the effects of GQM$^+$Strategies in this study. The audio clips and notes served as a reference when revising the grid: points of discussion that were unclear where checked against such documentation before any adjustments to the grid were made. Some qualitative data is used as secondary data to support the analysis of the quantitative data by providing explanations of specific observations.

## 4.3 Definitions

In order to assess the effects of GQM$^+$Strategies against data derived from each revision of the grid, we developed a goal classification scheme that helps to describe the different ways a grid can evolve over time. Based on this, we defined the following terms:

- A *new goal* is a goal that is introduced into the GQM$^+$Strategies grid and

that was not previously part of it.

- A *discarded goal* is a goal that has been removed from the grid.
- An *unchanged goal* is a goal that still exists in the grid without any changes compared to the previous version of the grid.
- A *revised goal* is a goal that has been altered in order to make corrections, refinements, improvements, or updates.
- *Split goals* are goals resulting from the separation of one goal into two or more goals. As a result, the original goals are discarded.
- *Merged goals* are goals that have been combined into a single goal.
- An *established goal* is a goal that has remained unchanged or been revised, split or merged compared to the previous grid revision.
- A *linked* goal is a goal that has been connected to a another goal in the GQM⁺Strategies grid.

### 4.4 Research Hypotheses

In order to analyse if our application of GQM⁺Strategies has indeed clarified and harmonized the goal set and aligned goals with strategies described in Section 2, we set the following hypotheses:

- **H1**: The percentage of new goals decreased over time during our application of GQM⁺Strategies.
- **H2**: The percentage of established goals increased over time.
- **H3**: The percentage of revised goals decreased over time.
- **H4**: The percentage of split goals decreased over time.
- **H5**: The amount of discarded goals decreases over time.
- **H6**: The percentage of linked goals increased over time.

We assume the following major links between the hypotheses and the expected benefits (see Section 2):

- **H1**, **H2** and **H5** indicate the fulfillment of benefit 2) (harmonization).
- **H3** and **H4** indicate the fulfillment of benefit 1) (clarity).
- **H6** indicates the fulfillment of benefit 4) (alignment of goals).

## 5 Results

Figure 1 presents the total number of goals and their transformations for each major GQM[+]Strategies grid revision. The total number of goals for each grid revision is shown in Table 2.

Figure 2 shows the evolution of new goals for each of our five major grid revisions. For the first grid revision, all 17 goals were new; for the last revision, the amount of new goals introduced had dropped to two. We interpret the figure as showing an overall decrease in new goals over time. Grid revisions 3–5, however, do not show a decreasing trend when examined independently of the two first revisions; specifically, the third revision introduces one new goal, the fourth revision no new goals and the third revision two new goals. We attribute this to the fact that the fourth workshop focused on identifying what historical data had already been collected by the organization—a thorough review of the grid under development was not the primary goal. Had it been the primary goal, grid revision 5 would likely not have seen two new goals; instead, these would have been defined as part of revision 4.

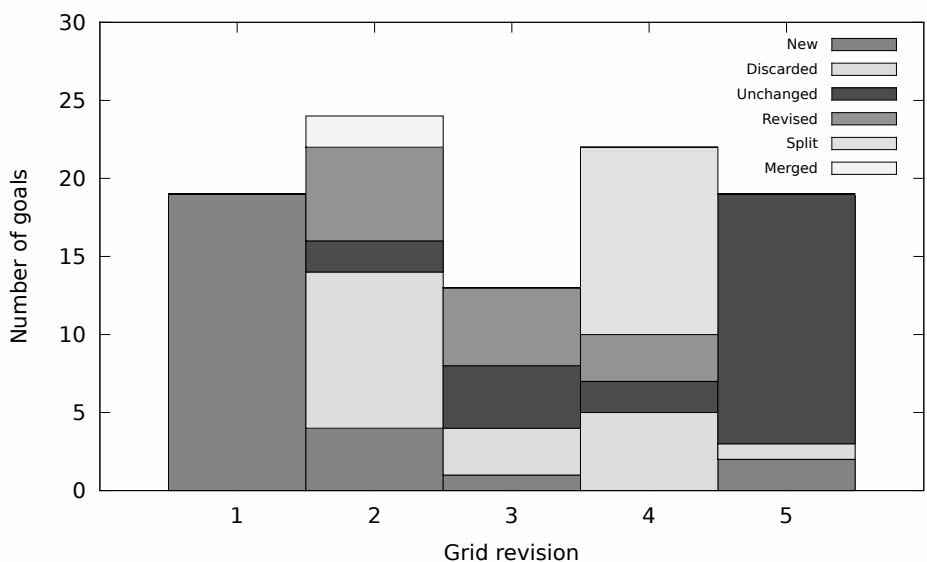

**Figure 1:** Goals and goal types per grid revision.

**Table 2:** Number of goals per revision.

| Revision | Total number of goals |
|---------:|----------------------:|
| 1 | 19 |
| 2 | 12 |
| 3 | 10 |
| 4 | 17 |
| 5 | 18 |

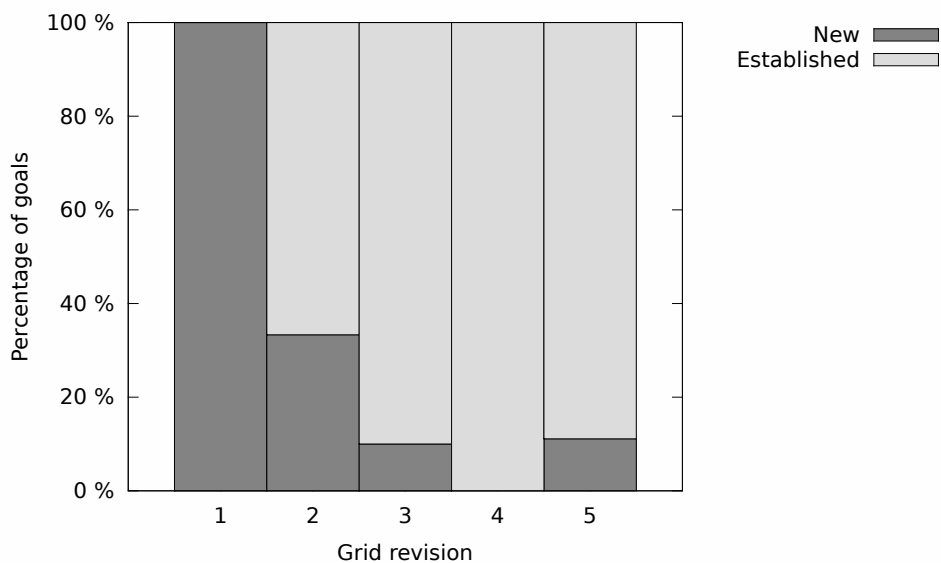

**Figure 2:** New goals distribution per grid revision.

The overall decreasing trend in new goals supports hypothesis H1. Conversely, we observe an increase in percentage of established goals over time, indicating that hypothesis H2 is also supported.

Figure 3 shows the evolution of revised goals for each of the five grid revisions. As changes to a goal can happen only after the first grid revision, the second revision is the first to include them (six in total). The percentage of revised goals in models 2 and 3 is the same, but the absolute values (five and three revisions, respectively) indicate a decreasing trend. We interpret this figure as showing an overall decrease in revised goals, supporting hypothesis H3.

Figure 4 details the evolution of split goals during each grid revision. The only revision where split goals were introduced was revision 4 (twelve split goals resulting from five previous goals). This was due to the revisiting of the organization's common hardware and software platform. In this case, split goals were so uncommon that we cannot assess hypothesis H4.

Figure 5 details the amount of goals discarded in each grid revision. A total of ten goals were discarded in the second revision, followed by three, five and one goal in subsequent revisions. The spike in discarded goals seen in revision 4 is a direct result of goal splitting, as observed in Figure 5. Overall, we observe a decreasing trend in discarded goals. Thus, hypothesis H5 is supported.

Figure 6 details how goals became linked over time. Two goals were linked in grid revision two, followed by nine, 17 and 18 goals. The significant increase in devised strategies in revision 3 can be attributed to the fact that workshop 3 was the first workshop in which strategies were discussed at length. Overall, the figure clearly shows the increase in linked goals over time; thus, hypothesis H6 is supported.

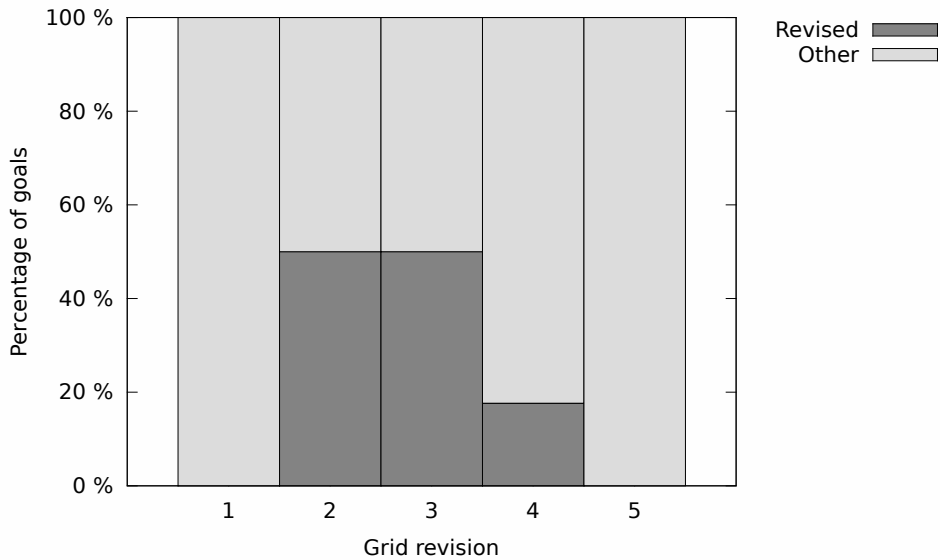

**Figure 3:** Revised goals distribution grid revision.

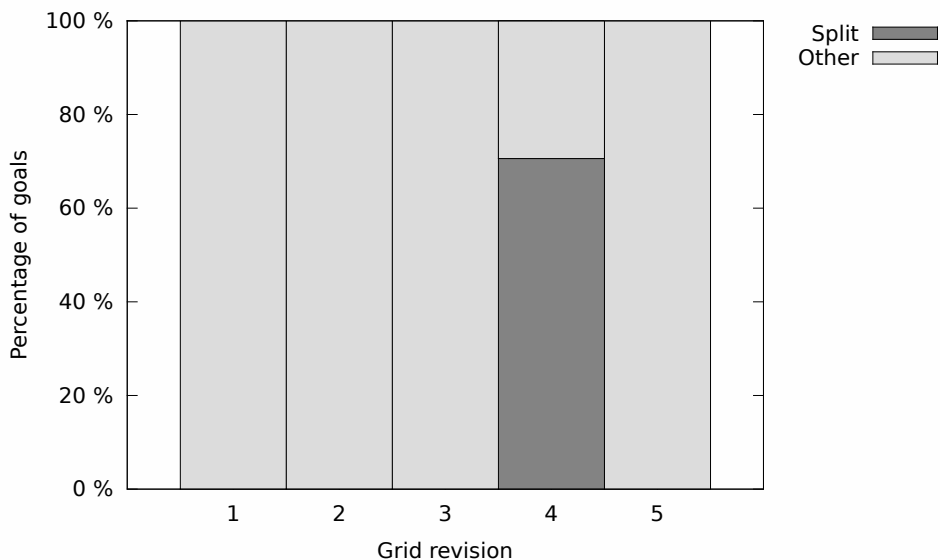

**Figure 4:** Split goals distribution per grid revision.

In all, hypotheses H1, H2, H3, H5 and H6 are supported for our application of GQM[+]Strategies. We can deduce that the approach has harmonized the goal set and successfully aligned goals. While hypothesis H4 could not be assessed, the stated benefit of clarity has been partially achieved through hypothesis H3.

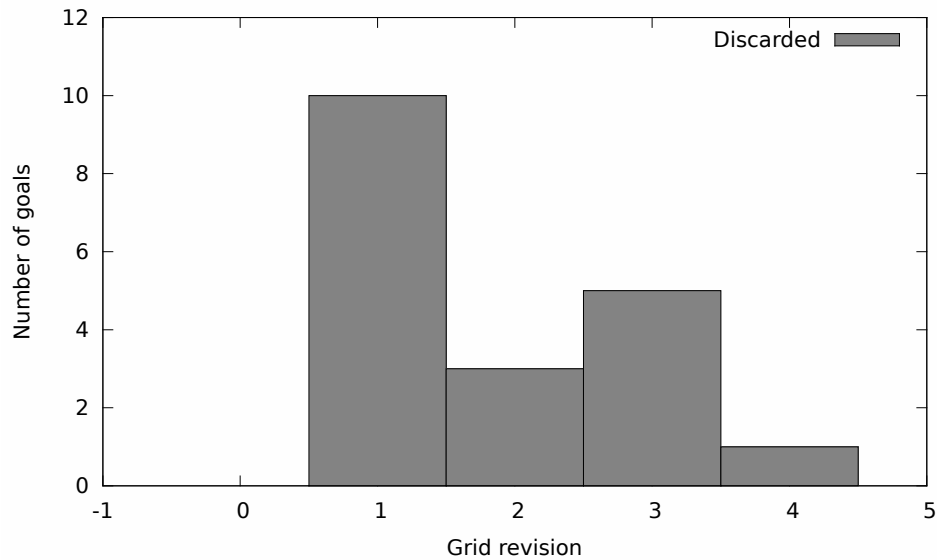

**Figure 5:** Discarded goals distribution per grid revision.

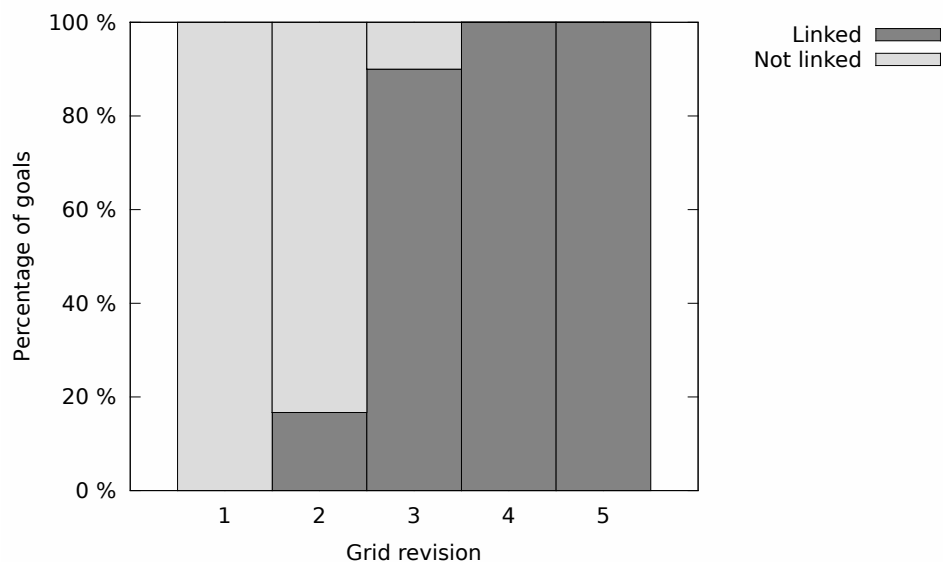

**Figure 6:** Linked goals distribution per grid revision.

## 6  Discussion and Limitations

All approaches to goal alignment strive towards the same end result: a set of strategic goals aligned in such a fashion that the satisfactory achievement of one goal has a positive impact on other goals. In a properly aligned organization, this linkage is consistent on all levels. Satisfaction of goals on each level contributes to the satisfaction of goals on higher levels, until ultimately a small set of top-level goals are satisfied. For the field of software engineering, this means sharing a common understanding of how software development can contribute to business goals and vice versa.

In our case study data, we traced the evolution of the GQM$^+$Strategies grid. After several rounds of refinement, the grid-under-development focused the attention on certain key goals, which were confirmed by the company stakeholders to be the ones that actually contribute to the business goals. We note that the resulting goal-set was significantly different from the initial set provided by the stakeholders, i.e., the process produced a better understanding for the company about its own goals.

The process of developing a GQM$^+$Strategies grid has two important outcomes: first, the grid itself is the basis for a streamlined measurement program that focuses on measuring achievement of goals that are actually relevant for organizational performance. Second, the grid development process itself is important for the organization to clarify, harmonize, and align its goals and strategies. However, developing the grid is not always straightforward. In particular, it may be challenging to track the evolution of the grid from one revision to the next. The goal classification scheme developed for this case study may assist in applying GQM$^+$Strategies in two ways: 1) it helps prioritize work on those goals which need expert input (new goals), final confirmation (revised goals), or which are potential candidates for linkage (split and merged goals), and 2) it allows monitoring the progress of grid creation and determining whether to stop introducing new goals and focusing on finalizing the grid. Thus the scheme may be used to assist in the iterative process of developing a GQM$^+$Strategies grid by parallelizing some of the activities according to goal class.

The results presented in this work are limited by a number of threats to validity. Being a single-case study, the results cannot be directly generalized to other contexts. In addition, following Yin [15], we discuss construct, internal, and external validity as well as reliability.

The construct validity in this study can be taken to mean the degree to which our measurements of alignment in the GQM$^+$Strategies grid actually measure what they claim to measure. Since we operate on the basic units of the grid using clearly defined operations, the construct validity of our measurements should be high.

Some confounding factors may be present that threaten internal validity. The results may have differed if the period between workshops with the target organization were scheduled at regular, shorter intervals. The time between workshops led researchers to take some initiative in revising the grid prior to consulting the target organization, thus possibly impacting the ways in which grid versions evolved. Changes were, however, always approved by organization representatives as soon as possible.

External validity reflects the degree to which findings can be generalized, and the extent of usefulness beyond the investigated case. Though our results indicate a fully aligned set of goals and strategies, a complete determination of the benefits and further impacts of such alignment entails implementing strategies and measuring their effects; such actions have yet to be taken. Longitudinal case studies would be needed to assess the degree of the final goals of achieving the targeted business objectives.

It should be noted that GQM$^+$Strategies grids need continuous evolution over time. Having a consolidated version of a does not mean that it is static. Due to changes in business or organizational context as well as for other reasons, grids should be continuously evaluated and evolved.

## 7 Conclusions and Future Work

In this paper, we sought to determine if the GQM$^+$Strategies approach provides a sound mechanism through which the alignment of goals on different levels of an organization can be achieved. The results of the case study indicate that for the purposes of clarifying, harmonizing and aligning goals with strategies, GQM$^+$Strategies is a beneficial approach.

Through the case study, we developed a goal classification scheme that may benefit future applications of GQM$^+$Strategies. Using the scheme, work on a GQM$^+$Strategies grid may be prioritized, monitored, and potentially parallelized. This could improve adoption of the approach in industry.

Future directions of this work involve both further development of the grid produced in the case company as well as potential tests with deployment in the test company. Mapping the elements of the produced GQM$^+$Strategies grid to Elektrobit's internal customer satisfaction measurement program is ongoing. By understanding how the defined goals and strategies relate in particular to customer satisfaction, we expect to recognize and measure the extent to which particular strategies have significant impact on customers' experiences throughout a development project.

### 7.0.1 Acknowledgments

This work was supported by TEKES as part of the Cloud Software program of DIGILE (Finnish Strategic Centre for Science, Technology and Innovation in the field of ICT and digital business).

*Jürgen Münch, Fabian Fagerholm, Petri Kettunen, Max Pagels, Jari Partanen*